\documentclass[eqno,pre,showpacs,reprint,twocolumn]{revtex4-1}

\usepackage{epsfig}
\usepackage{amsfonts}
\usepackage{amssymb}

\newcommand{\ak}{\hspace{20pt}}

\newcommand{\ms}{\medskip}
\newcommand{\bs}{\bigskip}

\def\bbbz{{\mathchoice {\hbox{$\sf\textstyle Z\kern-0.4em Z$}}
{\hbox{$\sf\textstyle Z\kern-0.4em Z$}} 
{\hbox{$\sf\scriptstyle Z\kern-0.3em Z$}} {\hbox{$\sf\scriptscriptstyle
Z\kern-0.2em Z$}}}}
\def\bbbI{{\rm 1\kern-0.25em I}}

\begin{document}

\title{
Some remarks about the domain statistics in the Ising model}
\author{K. Lukierska-Walasek}
\email{k.lukierska@uksw.edu.pl}\affiliation{Faculty of Mathematics and Natural Sciences, Cardinal Stefan Wyszynski Uniwersity, Dewajtis5, 501-815 Warsaw, Poland}
\author{K. Topolski}
\email{topolski@math.uni.wroc.pl}\affiliation{Institute of Mathematics, Wroc\l aw University, Pl. Grunwaldzki 2/4, 50-384 Wroc\l aw, Poland}
\author{M.Wodawski}\affiliation{Faculty of Mathematics and Natural Sciences, Cardinal Stefan Wyszynski University, Dewajtis 5, 01-815 Warsaw, Poland}

\date{\today}
\
\begin{abstract}
In this paper we describe a relation between the Zipf law and  statistical distributions for the Fortuin-Kasteleyn clusters in the Ising model.It has been shown,that histograms for fixed domain masses  present the right-skewed distributions.

\end{abstract}
\maketitle
\noindent

In our paper we consider  power law distributions, especially the Pareto distributions. These distributions have been investigated for the  last few decades in  connection with various subjects for examples  languages \cite{Zip}, analysis of populations in cities\cite{Bat},the turnover of Europes's largest companies\cite{Bou}, biology\cite{Wuc} and  physics\cite{Ma},\cite{Sci},\cite{Car},\cite{Klt}, \cite{Klt2}.\\
Although in this paper we consider a problem rather  well understood qualitatively, it is still far from being closed and it requires some explanation.
We study the size distributions of the Fortuin-Kasteleyn( F-K)clusters obtained by the Monte Carlo simulations in the two- dimensional Ising model. It is well known,these clusters obey the power law at criticality. In our case the measure of size is the number of spins in the cluster ( the mass of the cluster), but it can be also another measure of  size as a disk covering cluster or  an area enclosed  by  boundaries \cite{Car}. The corresponding fractal exponent may be related to the criticality exponent as is shown in\cite{Cog},or more recently in \cite{Jan}. Similar situation we have for a percolation.This idea for the Ising model in 2d may be found in\cite{Ono}.\\
The idea of the power law is explaned as consequence of the scale invariance at a criticality due to the divergence of the correlation length . Away from the criticality, the range of lengths, where the system is fractal, is limited by the finite correlation lenght, so that the power-law tail of the distribution is suppressed.\\

\ak 
We consider the Hamiltonian for the Ising model :

\vspace{-3mm}
\begin{equation}
H=-\frac{1}{2}\sum_{ij} J_{ij} \;S_i^z \;S_j^z\
\end{equation}

\ms
with a sum over all neighbouring pairs (z-$th$ components) of spins. Usually it is
assumed that the crystal lattice of  a ferromagnet is regular and in each site of
a lattice the spin is localized with the value $\,S^z=1\,$ or $\,S^z=-1$.
Further we have:

\begin{equation}
J_{ij}=\left\{
\begin{array}{l}
J \quad \mbox{if \ $i,j$ are neighbouring pairs of spins}\\[2mm]
0 \quad \mbox{ if \ 
$i,j$ are opposite each other}.
\end{array}\right\}
\end{equation}

\bs
Two spins $\,i\,$ and $\,j\,$ interact with each other by an energy $\ \pm J,
S_i^z\,S_k^z\,$ with $-J\,$ if both spins are parallel and $+J\,$ if
they are opposite each other. The energy needed for flipping of one spin is
$\,2J$.

Results presented hereafter are  obtained by applying the Monte Carlo techniques, 
based on the Swendsen-Wang cluster algorithm \cite{Sww} to the two-dimensional Ising model  
with periodic boundary conditions. 
In this algorithm clusters of spins are created by  introducing bonds between
neighboring spins of the same orientation, with the 
probability $1-\exp(-\frac{\Delta E}{k_B T}),\,$
where $\,k_B\,$ is the Boltzmann constant and  $\,\Delta E\,$
is the energy required to transform a pair of equal spins to a pair
of opposite spins. To each cluster we assign at random $-1$ or $+1$ orientation.\\
The different clusters are getting independent orientations.
Starting with the simulation having a random distribution
of a half of the spins up and a half of the spins down  and using the Swendsen-Wang
algorithm with a low temperature one sees the growing domains, in which spins are parallel.
We have two kinds of domains: with spin up and with spins down.
 
The phase transition appears at the critical point $\,T=T_c\,$ 
($\,T_c=2J/k_B\,\ln(1+\sqrt{2})\,$).
The difference  between the number of spins up and down $\,M\,$ is proportional
to the magnetization and near the critical point vanishes as $\,(T-T_c)^\beta$,
where for the dimension $\,d=2$, $\,\beta=\frac{1}{8}$. The correlation length
$\,\xi\sim|T-T_c|^\nu$, while the magnetization is proportional to $\,\xi^{-\beta/\nu}$.
In a finite system at the critical temperature $T_c$ one can replace $\,\xi$ by $L$,
hence $\,M\sim L^{d-\beta/\nu}=L^D\,$ with the fractal dimension
$\,D=d-\beta/\nu \ \ (d\leq 4)$.\\
In this paper we want to show that the right-skewed distributions of the clusters  are present  in the histograms with the fixed  masses and we explain the relation between the phase transition , the Zipf law ,the power law and the Pareto distributions.\\
Firstly  in this paper we want to underline that the Zipf law denotes greater restriction than the existence of the power law expansion.
The power law is presented by the Mandelbrot law in the form \cite{Man}:
\begin {equation}
x = C/( k + \alpha)^{1/\mu}
\end{equation}
where C and $\alpha$  and $\mu$ are some constants and $x$ denotes  the value of the object, $k$ is the rank order of the object.The object with the largest variable value is ranked as the first, the next with a smaller value is ranked as the second and so on.In this way we obtain the rank order $k$ of the object.
 The Mandelbrot law  undergoes  the Zipf law when $ \alpha = 0 $ and $\mu = 1$ :
\begin{equation}
 x = C/k
\end{equation}
 
\begin{figure}[h]
\centerline{\includegraphics[width=3.5in]{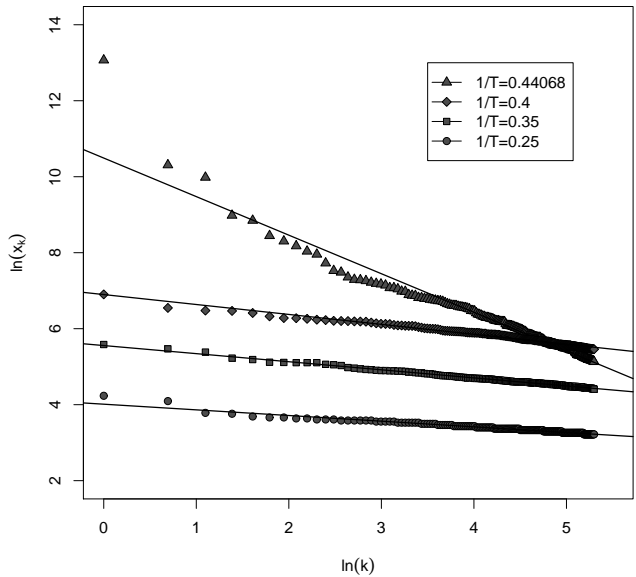}}
\caption{\label{Fig1} The log-log distribution of the domain masses $x_k$ versus the rank order index $k$   for  the two-dimensional Ising model for $ T=2.2692$(the critical temperature) ($\Diamond$),$T=2.5000$ ($\ast$), $T=2.8571$ ($\Box$) and $T=4.0000$ ($\circ$) for  $L=1000$.}
\end{figure}

This equation performs a straight line in a double logarythmic plot with the slope equal to $\pi /4 $.\\
In the case of the Ising model $x$ denotes the domain mass of the rank $k$(see Fig.\ \ref{Fig1}). 

The straight lines  with the slopes near $-1/\mu$ are given in Table \ref{Tab1}, 
 correspond to temperatures greater than the critical temperature $T_c$ ($T_c = 2.2692$).
 
\begin{table}[h]
\caption{\label{Tab1}The estimation of indexes $\mu$ with standard deviations $s$.} 
\begin{center}
\begin{tabular}{|l|l|l|}
\hline
$ T$ & $\mu $ & $s$ \\ \hline
$4.0000$ & $  6.7136$ & $0.0477$ \\
$2.8571$ & $ 4.7018$ &$0.0603$\\ 
$2.5000$ & $3.8240$ & $0.0695$\\  
$T_{c}\simeq 2.2692,\quad $ & $0.9854$ & $0.0874$\\ 
\hline
\end{tabular}
\end{center}
\end{table}

The estimations of the indexes $\mu $ for the two-dimensional Ising model with different temperatures based on 1000 realisations for $L=1000$ lattice are presented in Table \ref{Tab1}.
 and Fig.\ \ref{Fig1}.\\
It is seen from Fig.\ \ref{Fig1}. that a slope $arctan(1/{\mu})$ of the line only for the critical temperature is near $\pi/4$ for the critical temperature  and the Zipf law is satisfied. For the temperatures above critical temperature, slopes are smaller than $\pi/4$ and the Zipf law is not valid.
One can notice that the validity of the Zipf law when the phase transition is present  is in agreement with the relation $ 1/\mu = |2H-1|$ \cite{Czi}, where $H$ is the Hurst exponent.It is well known that the system exhibits infinite long correlations at the critical point,which corresponds to the biggest value of the Hurst exponent.In such a case one obtains $\mu =1$.\\
The Pareto distribution implies the Zipf law  \cite{Tro},\cite{Klt},but not inversally - for instance the Zipf law is satysfied also in the case of the hyperbolic distributions\cite{Har}.\\
The probability density of the Pareto distributions is defined as follows:
\begin{equation}
\rho(x) = (\mu/x_{0})(x{_0}/x)^{\mu+1}
\end{equation}

for $x > x_0$, where $x_0$ denotes a typical scale.

One can define so called $\mu$-variable\cite{Bou},\cite{Bou1}.The main property of $[\mu]$-variable is that all its moments $ m_q = < X^q > $ with $q \geq \mu $ are infinite.
The question in  classic probabilistic theory answered by L\'evy and Khintchine
shows how to characterize the limit distribution of the sum of N independent random variables $x$.
Suppose that $\rho(x)$ decreases for a large x such as $ x^{-1-\mu}$, then considering power-law distributions we can distinquish the following ranges for the critical exponents $\mu$ :\\

(i)For $\mu > 2$, the moment of the first order  $m_1$  and the moment of the second order $m_2$ are finite. \\ 
For this range of $\mu$,the two parameters $m_1$ and $m_2$ are finite and for large N the central limit theorem applies and the statistical distribution of the system tends towards a Gaussian distribution.\\
( ii)For $1< \mu < 2$ only the moment of the first order is finite  but the moment of the second order is infinite.\\
(iii) For $ \mu < 1$ the moments of the first  and the second order are infinite.For a large number of elements of the system it is described according to the L\'evy stable distribution, for instance the Pareto distribution.\\
But magnetic domains  described by the clusters of the Ising or the Potts models are not  independent variabes. Although we do not have  a correspondence between $\mu$ and the slope of distributions described in (i)-(iii), respectively, the presence of the Pareto-type  dystrybutions are also possible in the case of the dependent  variables .
\bigskip
\begin{figure}[h]
\centerline{\includegraphics[width=3.5in]{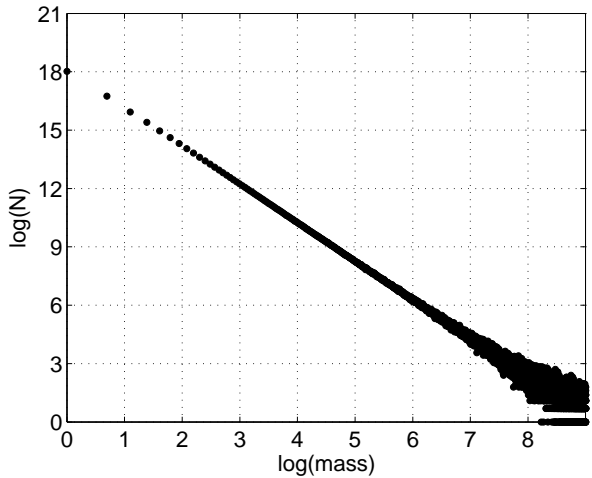}}
\caption{\label{Fig2} The dependence between the  number of domains with mass x normalizedby the number of configurations $(N_x/N)$ from the mass $(x)$in the log-log scale, for the two dimensional Ising model near the criticalpoint $T_c$, for $10000$ configurations and  $L=600$.}
\end{figure}

The straight line  of Fig.\ \ref{Fig2} presents the Pareto distribution.

In our case, away from the criticality, the Gaussian-type distributions are not present when $\mu > 2$. 
In this region the range of lengths, where the system is fractal, it is limited by the finite correlation length and the power-law tail in the  distribution is suppressed. 

\begin{figure}[h]
\centerline{\includegraphics [width=4in]{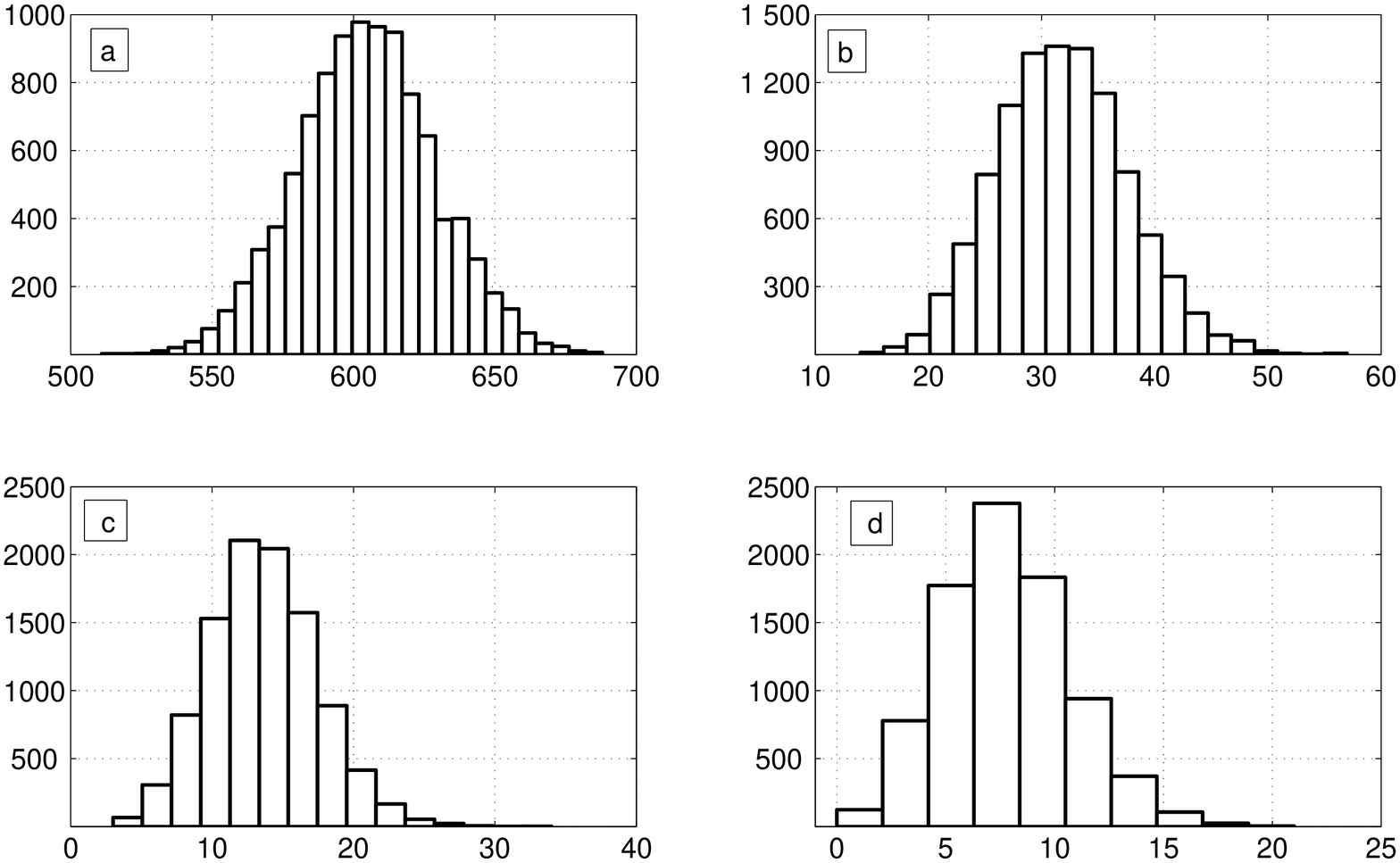}}
\caption{\label{Fig3} Histogram of the fixed domain mass $ m=5$(a),$m=10$(b), $m=50$(c) and $m=70$(d) for the inverse temperature $\beta=025$, for the system with $L=600$, for $10000$ configurations.}
\end{figure}

Fig.\ \ref{Fig3} presents the distributions of the number of configurations with defined mass.The total number of configurations is $10000$.Histograms are presented for masses $x=5$, $x=10$, $x=50$ and$ x=70$ for , for a system with $L=600$, beyond the phase transition inverse  temperature $\beta = 0.25$.\\

 The test for the Gaussian distribution was performed using the Kolmogorow-Smirnov method and the chi-square test. The Gaussian distributions for the number of clusters of a fixed size are present only for the masses smaller than 5. For example for the temperature $T= 4.000$ and the cluster of size 5 the p-value of the chi-squared test was equal to$ 0.0771$.For the clusters of larger size we obtain the right-skewed distributions.

\begin{figure}[h]
\centerline{\includegraphics[width=4in]{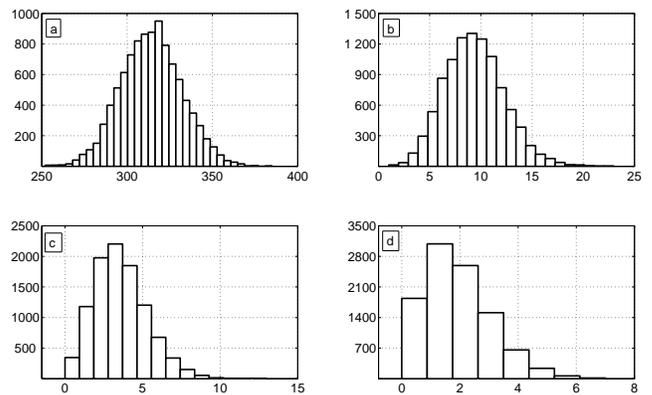}}
\caption{\label{Fig4}Histogram of the fixed mass $ m=5$(a), $m=10$(b), $m=50$(c) and $m=70$(d) for  the critical temperature,  for the system with $L=600$, for 10000 configurations.}
\end{figure}

Fig.\ \ref{Fig4} presents the distribution of the number of configurations with a defined mass.The total number of configurations is $10000$.Histograms are presented for masses $x=5$, $x=10$, $x=50$ and
$ x=70$ for ,for a system with $L=600$ near critical point ($T_{c}=2.2692$).
 
In this case the right-skewed distributions are present.The Gaussian distributions we observe only for masses smaller than 3. For example for the cluster of size 5 the p-value of the chi-square test was equal to 0.01174.

The main conclusion from this paper is following: In the critical region  the Zipf law is satisfied and the power law distributions, the Pareto distributions are present.The histograms of the numbers od domain configurations,which possess the same number of domains present the right-skewed distributions mainly, for the critical point as well beyond it. The Gaussian-type distributions appear only for small masses . 
 


\end{document}